\documentclass[useAMS,usenatbib]{mn2e}
\usepackage{url}  

\usepackage{graphicx}

\newcommand       \be           {\begin{equation}}
\newcommand       \ee           {\end{equation}}
\newcommand\aj{ {AJ}}% Astronomical Journal 
\newcommand\apj{ {ApJ}}% Astrophysical Journal 
\newcommand\apjl{ {ApJ}}% Astrophysical Journal, Letters 
\newcommand\apjs{ {ApJS}}% Astrophysical Journal, Supplement 
\newcommand\aap{ {A\&A}}% Astronomy and Astrophysics 
\newcommand\aapr{ {A\&A~Rev.}}% Astronomy and Astrophysics Reviews 
\newcommand\mnras{ {MNRAS}}% Monthly Notices of the RAS 
\newcommand\pasp{ {PASP}}% Publications of the ASP 
\newcommand\nat{ {Nature}}% Nature 
\newcommand\physrep{ {Phys.~Rep.}}% Physics Reports 
\def\simlt{\mathrel{\hbox{\rlap{\hbox{\lower4pt\hbox{$\sim$}}}\hbox{$<$}}}}
\def\simgt{\mathrel{\hbox{\rlap{\hbox{\lower4pt\hbox{$\sim$}}}\hbox{$>$}}}}

\def\simlt{\mathrel{\hbox{\rlap{\hbox{\lower4pt\hbox{$\sim$}}}\hbox{$<$}}}}
\def\simgt{\mathrel{\hbox{\rlap{\hbox{\lower4pt\hbox{$\sim$}}}\hbox{$>$}}}}
\def\lesssim{\mathrel{\hbox{\rlap{\hbox{\lower4pt\hbox{$\sim$}}}\hbox{$<$}}}}

\def\gtrsim{\mathrel{\hbox{\rlap{\hbox{\lower4pt\hbox{$\sim$}}}\hbox{$>$}}}}

\title[Radioactively Powered Transients from Compact Object Mergers]{Electromagnetic counterparts of compact object mergers powered by the radioactive decay of $r$-process nuclei} \author[Metzger et al.]{B.~D. Metzger$^{1,11}$\thanks{E-mail: bmetzger@astro.princeton.edu}, G.~Mart{\'{\i}}nez-Pinedo$^{2}$, S.~Darbha$^{3}$, E.~Quataert$^{3}$, A.~Arcones$^{2,4}$, 
\vspace{0.25cm}
\\{\LARGE\rm D.~Kasen$^{5,12}$, R.~Thomas$^{6}$, P.~Nugent$^{6}$, I.~V.~Panov$^{7,8,9}$, $\&$ N.~T.~Zinner$^{10}$} \\ $^{1}$ Department of Astrophysical Sciences, Princeton University, Princeton, NJ 08544, USA \\ $^{2}$ GSI Helmholtzzentrum f\"{u}r Schwerionenforschung,
Planckstr.~1, D-64291 Darmstadt, Germany\\ $^{3}$ Astronomy Department and Theoretical Astrophysics Center, University of California, Berkeley, 601 Campbell Hall, Berkeley CA, 94720 \\ $^{4}$ Institut f\"{u}r Kernphysik, TU~Darmstadt,
Schlossgartenstr.~9, D-64289 Darmstadt, Germany \\ $^{5}$ University of California, Santa Cruz, CA 95064, USA; \\ $^{6}$ Computational Cosmology Center, Lawrence Berkeley National Laboratory, 1
Cyclotron Road MS50B-4206, Berkeley, CA, 94720 \\ $^{7}$ Department of Physics, University of Basel, Klingelbergstr.~82, CH-4056 Basel, Switzerland \\
$^{8}$ Institute for Theoretical and Experimental Physics, B.~Cheremushkinskaya St.~25, 117259, Moscow, Russia \\
$^{9}$ Russian Research Centre Kurchatov Institute, pl.~Kurchatova 1, Moscow, 123182, Russia \\
$^{10}$ Department of Physics, Harvard University, Cambridge, MA 02138, USA \\
 $^{11}$ NASA Einstein Fellow \\
$^{12}$ Hubble Fellow}
\begin{document}
\date{Accepted . Received ; in original form }
\pagerange{\pageref{firstpage}--\pageref{lastpage}} \pubyear{????}
\maketitle
\label{firstpage}

\begin{abstract}

The most promising astrophysical sources of kHz gravitational waves (GWs) are the inspiral and merger of binary neutron star(NS)/black hole systems.  Maximizing the scientific return of a GW detection will require identifying a coincident electro-magnetic (EM) counterpart.  One of the most likely sources of {\it isotropic} EM emission from compact object mergers is a supernova-like transient powered by the radioactive decay of heavy elements synthesized in ejecta from the merger.  We present the first calculations of the optical transients from compact object mergers that self-consistently determine the radioactive heating by means of a nuclear reaction network; using this heating rate, we model the light curve with a one dimensional Monte Carlo radiation transfer calculation.  For an ejecta mass $\sim 10^{-2}M_{\sun}$[$10^{-3}M_{\sun}$] the resulting light curve peaks on a timescale $\sim 1$ day at a V-band luminosity $\nu L_{\nu} \sim 3\times 10^{41}$[$10^{41}$] ergs s$^{-1}$ ($M_{\rm V}$ = $-$15[$-$14]); this corresponds to an effective ``f'' parameter $\sim 3\times 10^{-6}$ in the Li-Paczynski toy model.  We argue that these results are relatively insensitive to uncertainties in the relevant nuclear physics and to the precise early-time dynamics and ejecta composition.  Since NS merger transients peak at a luminosity that is a factor $\sim 10^{3}$ higher than a typical nova, we propose naming these events ``kilo-novae.''  Due to the rapid evolution and low luminosity of NS merger transients, EM counterpart searches triggered by GW detections will require close collaboration between the GW and astronomical communities.  NS merger transients may also be detectable following a short-duration Gamma-Ray Burst or ``blindly'' with present or upcoming optical transient surveys.  Because the emission produced by NS merger ejecta is powered by the formation of rare $r$-process elements, current optical transient surveys can directly constrain the unknown origin of the heaviest elements in the Universe.

\end{abstract}


\vspace{-0.7cm}
\section{Introduction}
The direct detection of gravitational waves (GWs) would be a major breakthrough for both fundamental physics and astrophysics.  With upgrades of the ground-based interferometers LIGO \citep{Abramovici:1992} and Virgo \citep[e.g.][]{Caron:1999} to ``advanced'' sensitivity 
expected within the next decade, GW detection is rapidly becoming a 
realistic$-$even anticipated$-$possibility.  

The most promising astrophysical sources of GWs for ground-based detectors are thought to be the GW-driven in-spiral and coalescence of binary compact objects (neutron stars [NSs] and black holes 
[BHs]).  Advances in general relativistic simulations of the merger process \citep{Pretorius:2005} are honing our understanding of the strength and form of the expected signal \citep[see e.g.][for recent reviews]{Faber+:2009,Duez:2009}.  However, estimates of the merger rates based on known NS-NS binaries and population synthesis remain uncertain by at least an order of magnitude \citep{Kim+:2005,Belczynski+:2006,Kalogera+:2007,LIGO+:2010}.  For instance, \citet{Kim+:2005} estimate that the NS-NS merger rate detectable with advanced LIGO will be $27^{+62}_{-21}$ per year, implying that if the true rate lies on the low end of present estimates then only a few sources may be detected per year.  This possibility makes it especially crucial that we extract the most science from each event.

Optimizing the science from a detected GW signal requires identifying a 
coincident electromagnetic (EM) counterpart \citep[e.g.][]{Schutz:1986,Schutz:2002,Sylvestre:2003,Stubbs:2008,Bloom+:2009,Phinney:2009,Stamatikos+:2009}.  By independently identifying the source's 
position and time, several of the degeneracies associated with the GW 
signal are lifted \citep{Hughes.Holz:2003,Arun+:2009} and the signal-to-noise 
required for a confident detection is decreased \citep{Kochanek.Piran:1993,Dalal+:2006}.  Coupled with its GW-measured luminosity distance, identifying the merger's redshift (e.g. by localizing its host galaxy) would also allow for a precision measurement of the Hubble constant \citep[e.g.][]{Krolak.Schutz:1987,Holz.Hughes:2005,Deffayet.Menou:2007}.  Likewise, the potential wealth of complementary information encoded in 
the EM signal may be essential to fully unraveling the astrophysical context 
of the event \citep{Phinney:2009}. 

The most commonly discussed EM signal associated with NS-NS/NS-BH mergers 
is a short-duration Gamma-Ray Burst (GRB), powered by the accretion of 
material that remains in a centrifugally-supported torus around the BH 
following the merger \citep{Paczynski:1986,Narayan+:1992}.  The {\it 
Swift} satellite \citep{Gehrels+:2004} has revolutionized our 
understanding of short GRBs by detecting and localizing a significant number of their afterglows 
for the first time.  This has enabled the discovery that short 
GRBs likely originate from a more evolved stellar population than those of long-duration GRBs (e.g. \citealt{Bloom+:2006}; \citealt{Berger+:2005}; \citealt{Villasenor+:2005}; see \citealt{Berger:2009} for a recent review), consistent with an origin associated with compact object mergers \citep{Nakar+:2006}.  Despite these suggestive hints, however, it is not yet established 
that all short GRBs are uniquely associated with NS-NS/NS-BH mergers (e.g.~\citealt{Hurley+:2005}; \citealt*{Metzger+:2008b}) nor that all mergers lead to an energetic GRB.  Furthermore, only a small fraction of GRB jets are pointed towards us \citep{Rhoads:1999} and for off-axis events, the prompt and 
afterglow emission are much dimmer due to relativistic de-beaming.  Although some emission may be observed by off-axis viewers, such ``orphan'' afterglows \citep[e.g.][]{Totani.Panaitescu:2002} are typically expected to peak at radio wavelengths on a timescale of months-years 
\citep{Soderberg+a:2006,Rossi+:2008,Zhang.MacFadyen:2009}.  Only a limited fraction of short GRBs are detected in radio, even when viewed on-axis \citep[e.g.][]{Soderberg+b:2006}.

In parallel to the advances in GW detectors, the advent of large-scale optical surveys with increasing sensitivity, rapid cadence, and sky-area coverage is leading to a revolution in the study of transient objects.  These include the 
Palomar Transient Factory \citep[PTF;][]{Rau+:2009}, Pan-STARRs \citep{Kaiser+:2002}, SkyMapper \citep{Keller+:2007}, and the VLT Survey Telescope \citep[VST; ][]{Mancini+:2000}, which are paving the way for future endeavors such as the Large Synoptic Survey Telescope \citep[LSST; ][]{Strauss+:2010} and the proposed Synoptic All Sky Infrared Imaging (SASIR) survey \citep{Bloom+b:2009}.  Given these present and anticipated future capabilities, the most promising EM counterpart of compact object mergers is arguably an {\it isotropic}, {\it optical/near infrared (NIR)} wavelength signal.  In addition to providing time-stamped maps of the night sky for use in follow-up observations, these ``blind'' surveys could also detect EM counterparts even independent of a GW or GRB trigger (see $\S\ref{sec:surveysearch}$).

One proposed source of relatively isotropic optical/NIR emission following a 
NS-NS/NS-BH merger is a supernova(SN)-like transient powered by the radioactive decay of merger ejecta (\citealt{Li.Paczynski:1998}; hereafter LP98; cf.~\citealt{Kulkarni:2005}; \citealt*{Metzger+:2008c}).  Although Type Ia supernova light curves are powered largely by the decay of $^{56}$Ni (e.g. \citealt{Kasen.Woosley:2009}), most of the ejecta from NS-NS/NS-BH mergers is highly neutron-rich (electron fraction $Y_{e} \sim 0.1-0.4$) and produces little Ni.  Instead, much heavier radioactive elements are formed via rapid neutron capture ($r$-process) nucleosynthesis following the decompression of the ejecta from nuclear densities (e.g. \citealt{Lattimer.Schramm:1974,Lattimer.Schramm:1976,Eichler+:1989,Freiburghaus+:1999}).  Although the $r-$process itself lasts only a matter of seconds, these newly-synthesized elements undergo nuclear fission, alpha and beta decays on much longer timescales as they descend to $\beta$-stability.  The resulting energy release can power detectable thermal emission once the ejecta expands sufficiently that photons can escape.  Due to the lower quantity of ejecta and its faster speed, however, the resulting transient is dimmer and evolves faster than a normal SN.  Transients from NS mergers are thus a challenge to detect and identify.

Although the basic LP98 model provides a qualitative picture of the thermal transients from NS-NS/NS-BH mergers, it makes a number of simplifying assumption and leaves several free parameters unconstrained, including the fraction of nuclear energy released and the precise distribution of decaying nuclei.  LP98 further assume that the photosphere radiates as a black body, which is a poor assumption at moderate optical depths and in light of the substantial UV line blanketing that may accompany the rich energy spectra of the very heavy nuclei that dominate the composition.  These details may be important for predicting the unique, ``smoking gun'' features of merger-related transients.  Because the transient sky is expected to be rich in its diversity (e.g. \citealt{Becker+:2004}), more detailed predictions may be essential to identifying candidate sources in real-time for deeper follow-up observations, especially considering the likelihood that only limited information (e.g. photometric colors) may be available.  Understanding the detailed {\it spectroscopic} properties of merger transients is clearly an important endeavor.  

In this work we present the first self-consistent calculations of
the optical/NIR counterparts to NS-NS/NS-BH mergers.  In particular,
our work goes beyond previous work in two important ways: (1) we use a
nuclear physics reaction network to calculate the radioactive heating
of the ejecta and (2) we employ the Monte Carlo radiative transfer
code SEDONA to more accurately model the light curve and colors of the
resulting EM transient.  We begin in $\S\ref{sec:prelim}$ with
preliminary considerations, including a discussion of the sources of
neutron-rich ejecta from NS-NS/NS-BH mergers
($\S\ref{sec:nrichejecta}$) and a brief review of the relevant
scalings for radioactively-powered transients
($\S\ref{sec:LPanalytics}$).  In $\S\ref{sec:heating}$ we describe the
nucleosynthesis that occurs as the material decompresses from nuclear
densities and our calculations of the resulting radioactive heating,
including a detailed discussion of the efficiency of
fission/$\beta-$decay thermalization ($\S\ref{sec:therm}$).  In
$\S\ref{sec:lightcurves}$ we present calculations of the light curves
and color evolution of NS-NS/NS-BH merger transients, highlighting the
unique features of these events and the primary uncertainties in the theoretical predictions.  We find that the peak luminosities of NS merger transients are typically $\sim$ few $\times 10^{41}$ ergs s$^{-1}$, or a factor $\sim 10^{3}$ larger than the Eddington luminosity for a solar mass object.  We therefore dub these events ``kilo-novae,'' since standard novae are approximately Eddington-limited events.   In $\S\ref{sec:detection}$ we discuss the implications of our results for the present constraints
on, and the future detection of, kilonovae from NS-NS/NS-BH mergers,
including the direct constraints that optical transient surveys place
on the astrophysical origin of $r$-process elements
($\S\ref{sec:rprocess}$).  We summarize our results and conclude in
$\S\ref{sec:conclusions}$.

\section{Preliminary Considerations}
\label{sec:prelim}

\subsection{Sources of Neutron-Rich Ejecta}
\label{sec:nrichejecta}

There are several potential sources of neutron-rich ejecta from
NS-NS/NS-BH mergers.  First, neutron-rich material can be ejected due
to tidal forces during the merger itself \citep{Lattimer.Schramm:1974,Rosswog+:1999,Rosswog:2005}.  The quantity of this {\it
  dynamically ejected} material depends sensitively on the NS-NS/NS-BH
binary parameters and the NS equation of state (e.g. \citealt{Rosswog:2005}).
Since this material primarily originates from the NS's neutron-rich
outer core, it has a typical electron fraction $Y_{e} \sim 0.03-0.1$
(e.g. \citealt{Haensel.Zdunik:1990a,Haensel.Zdunik:1990b,Rosswog:2005}).  The electron fraction probably
remains low since the ejecta remains cold (and hence thermal weak
interactions remain slow) due to adiabatic losses as the material
rapidly expands from nuclear densities (e.g. \citealt{Ruffert+:1997}; \citealt{Duez+:2009}).  A typical outflow speed is $v \sim 0.1$ c.

Neutron-rich material also originates from outflows from the accretion
disk on longer, viscous timescales.  Neutrino-heated winds are driven
from the disk for a variety of accretion rates and disk radii during
its early evolution (e.g.~\citealt*{Metzger+:2008c}; \citealt{Surman+:2008}; \citealt{Dessart+:2009}).  Although these outflows are generally
neutron-rich, they can be proton-rich in some cases (e.g. \citealt*{Metzger+:2008a}; \citealt{Barzilay.Levinson:2008}).  An even larger quantity of
mass loss occurs at later times once neutrino cooling shuts off, due
to powerful outflows driven by viscous heating and the nuclear
recombination of free nuclei into $\alpha-$particles (\citealt*{Metzger+:2008c}; \citealt{Metzger+:2009a}; \citealt{Lee+:2009}).  \citet{Metzger+:2009a} show that $\sim 20-50\%$ of the initial disk mass is ejected
with a range of electron fractions $Y_{e} \sim 0.1-0.4$.  The wind's
asymptotic speed in this case is also $v \sim 0.1-0.2$ c, set by the
$\sim 8$ MeV per nucleon released as heavy elements are formed.

In summary, considering both the tidally- and wind-driven ejecta from
NS-NS/NS-BH mergers, from $M_{\rm ej} \sim 0$ up to $\sim 0.1 M_{\sun}$ in neutron-rich ejecta is expected with $v \sim
0.1$ c and $Y_{e} \lesssim 0.2$, i.e. sufficiently neutron-rich to
undergo a robust (low entropy) third-peak $r$-process during its
subsequent expansion (e.g. \citealt{Hoffman+:1997}).  A similar amount of
material may be ejected with $Y_{e} \sim 0.2-0.4$.  Although this
material is not sufficiently neutron-rich to reach the third
$r$-process peak, it also produces heavy elements that contribute a
comparable radioactive heating rate (Fig.~\ref{fig:edot2}).

\subsection{Analytic Estimates}
\label{sec:LPanalytics}

The majority of the energy released by the $r$-process occurs on a
timescale of $\sim$ seconds (e.g. \citealt{Freiburghaus+:1999}; \citealt{Goriely+:2005}; \citealt{Metzger+:2010}).  However, most of this initial heating
(and any residual heat from the merger itself) is lost to adiabatic
expansion because the outflow is highly optically thick at these early
times.  A significant EM luminosity is only possible once the density
decreases sufficiently that photons can escape the ejecta on the
expansion timescale \citep{Arnett:1982}.  The photon diffusion time through
the outflow at radius $R$ is approximately 
\be 
t_{\rm d} =
\frac{B\kappa M_{\rm ej}}{c R},
\label{eq:tdiff}
\ee 
where $\kappa$ is the opacity and $B \simeq 0.07$ for a spherical
outflow (e.g. \citealt{Padmanabhan:2000}).  Setting this equal to the expansion
time $t_{\rm exp} = R/v$ gives the characteristic radius for the EM
emission to peak
\begin{eqnarray}
R_{\rm peak} &\simeq& \left(\frac{Bv\kappa M_{\rm ej}}{c}\right)^{1/2} \nonumber \\ &\approx& 1.2\times 10^{14}{\rm\, cm}\left(\frac{v}{0.1 c}\right)^{1/2}\left(\frac{M_{\rm ej}}{10^{-2}M_{\sun}}\right)^{1/2},
\label{eq:rpeak}
\end{eqnarray}
where we have taken $\kappa = 0.1$ cm$^{2}$ g$^{-1}$ as an estimate of the line opacity of the $r$-process ejecta, assuming it is similar to that of Fe-peak elements \citep{Pinto.Eastman:2000}.  We discuss the validity of this assumption further in $\S\ref{sec:sedona}$.  Assuming free expansion $R = vt$, $R_{\rm peak}$ is reached on a timescale \citep{Arnett:1982}
\be
t_{\rm peak} \approx
0.5{\rm\, days}\left(\frac{v}{0.1 c}\right)^{-1/2}\left(\frac{M_{\rm ej}}{10^{-2}M_{\sun}}\right)^{1/2}.
\label{eq:tpeak}
\ee 
The above expression is strictly valid only if $t_{\rm peak}$ exceeds the
  intrinsic radioactive decay lifetime of the ejecta. This condition is generally satisfied for $r-$process ejecta due to their rather short $\beta$-decay half-lives.  This short timescale $t_{\rm peak} \sim 1$ day compared to that of
a normal SN ($t_{\rm peak} \sim$ weeks) is one of the defining
characteristics of kilonovae from NS mergers.

Provided that the radioactive power can be approximated as a
decreasing power-law in time $\dot{Q} \propto t^{-\alpha}$ with
$\alpha < 2$, the brightness of the event depends most sensitively on
the amount of radioactive heating that occurs around the timescale
$t_{\rm peak}$: $Q_{\rm peak} = \int_{t_{\rm peak}}\dot{Q}dt \approx
\dot{Q}(t_{\rm peak})t_{\rm peak} = fM_{\rm ej}c^{2}$, where $f \ll 1$
is a dimensionless number (LP98).  Parametrized thus, the peak
bolometric luminosity is approximately
\begin{eqnarray}
&&L_{\rm peak} \simeq \frac{Q_{\rm peak}}{t_{\rm d}(R_{\rm peak})} \nonumber \\
&&\approx 5\times 10^{41}{\rm ergs\,\,s^{-1}}\left(\frac{f}{10^{-6}}\right)\left(\frac{v}{0.1 c}\right)^{1/2}\left(\frac{M_{\rm ej}}{10^{-2}M_{\sun}}\right)^{1/2},\nonumber \\
\label{eq:lpeak}
\end{eqnarray}
and the effective temperature is given by
\begin{eqnarray}
&&T_{\rm peak} \simeq \left(\frac{L_{\rm peak}}{4\pi R_{\rm peak}^{2}\sigma}\right)^{1/4} \nonumber \\
&&\approx 1.4\times 10^{4}{\,K}\left(\frac{f}{10^{-6}}\right)^{1/4}\left(\frac{v}{0.1 c}\right)^{-1/8}\left(\frac{M_{\rm ej}}{10^{-2}M_{\sun}}\right)^{-1/8}\nonumber \\
\label{eq:temppeak}
\end{eqnarray}

%%%%%%%%%%%%%%%%%%%%%%%%%%%%%%%%%%%%%% FIG 3 %%%%%%%%%%%%%%%%%%%%%%%%%%%%%%
\begin{figure}
\resizebox{\hsize}{!}{\includegraphics[]{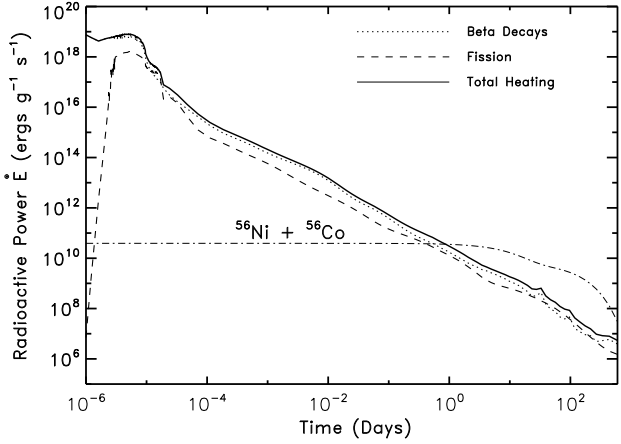}}
%\resizebox{\hsize}{5cm}{\includegraphics{1a.eps}}
%\includegraphics{1a.eps}
\caption{Radioactive heating rate per unit mass $\dot{E}$ in NS merger ejecta due to the decay of $r$-process material, calculated for the $Y_{e} = 0.1$ ejecta trajectory from Rosswog et al.~(1999) and Freiburghaus et al.~(1999).  The total heating rate is shown with a solid line and is divided into contributions from $\beta-$decays ({\it dotted line}) and fission ({\it dashed line}).  For comparison we also show the heating rate per unit mass produced by the decay chain $^{56}$Ni $\rightarrow$ $^{56}$Co $\rightarrow$ $^{56}$Fe ({\it dot-dashed line}).  Note that on the $\sim$ day timescales of interest for merger transients ($t \sim t_{\rm peak}$; eq.~[\ref{eq:tpeak}]) fission and $\beta-$decays make similar contributions to the total $r$-process heating, and that the $r$-process and $^{56}$Ni heating rates are similar.}
\label{fig:edot}
\end{figure}
%%%%%%%%%%%%%%%%%%%%%%%%%%%%%%%%%%%%%%%%%%%%%%%%%%%%%%%%%%%%%%%%%%%%%%%%%%%%
%%%%%%%%%%%%%%%%%%%%%%%%%%%%%%%%%%%%%%%%%%%%%%%%%%%%%%%%%%%%%%%%%%%%%%%%%%%%

%%%%%%%%%%%%%%%%%%%%%%%%%%%%%%%%%%%%%% FIG 3 %%%%%%%%%%%%%%%%%%%%%%%%%%%%%%
\begin{figure}
\resizebox{\hsize}{!}{\includegraphics[]{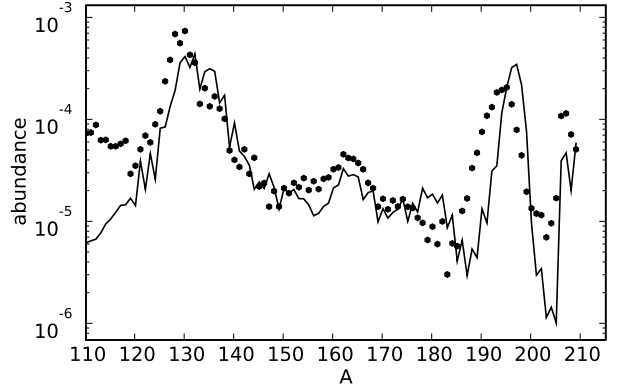}}
\caption{Final abundance distribution from the fiducial model with $Y_{e} = 0.1$ (Fig.~\ref{fig:edot}), shown as the mass fraction versus mass number $A$.  Measured solar system $r$-process abundances are shown for comparison with black dots.  They are arbitrarily normalized to the computed abundances for $A = 195$.}
\label{fig:abundances}
\end{figure}
%%%%%%%%%%%%%%%%%%%%%%%%%%%%%%%%%%%%%%%%%%%%%%%%%%%%%%%%%%%%%%%%%%%%%%%%%%%%
%%%%%%%%%%%%%%%%%%%%%%%%%%%%%%%%%%%%%%%%%%%%%%%%%%%%%%%%%%%%%%%%%%%%%%%%%%%%

Note that $L_{\rm peak} \propto f$, yet the value of $f$ is left as a free parameter in the LP98 model, with values up to $f \sim 10^{-3}$ considered plausible {\it a priori}.  In $\S\ref{sec:therm}$ we present explicit calculations of $\dot{Q}$ and show that the effective value of $f$ is $\sim 3\times 10^{-6}$.  Thus, for $M_{\rm ej} \sim 10^{-2}M_{\sun}$ we expect a transient with peak luminosity $\sim 10^{42}$ ergs s$^{-1}$ (bolometric magnitude $M_{\rm bol} \approx -16$) and a photospheric temperature $\sim 10^{4}$ K, corresponding to a spectral peak at optical/near-UV wavelengths.  

\section{Radioactive Heating}
\label{sec:heating}

\subsection{Network Calculations}
\label{sec:network}

In this section we present calculations of the radioactive heating of
the ejecta.  We use a dynamical $r$-process network (\citealt{Martinez-Pinedo:2008};\citealt{Petermann+:2008}) that includes neutron captures,
photodissociations, $\beta-$decays, $\alpha-$decays and fission reactions.  The latter
includes contributions from neutron induced fission, $\beta$ delayed
fission, and spontaneous fission.  The neutron capture rates for nuclei with $Z\leq83$ are
  obtained from the work of~\citet{Rauscher.Thielemann:2000} and are
  based on two different nuclear mass models: the Finite Range Droplet
  Model~\citep{Moeller.Nix.ea:1995} and the Quenched version of the
  Extended Thomas Fermi with Strutinsky Integral model
  (ETFSI-Q)~\citep{Pearson.Nayak.Goriely:1996}.  For nuclei with
  $Z>83$ the neutron capture rates and neutron-induced fission rates
  are obtained from~\citet{Panov.Korneev.ea:2009}. Beta-decay rates
  including emission of up to 3 neutrons after beta decay are
  from~\citet{Moeller.Pfeiffer.Kratz:2003}. Beta-delayed fission and
  spontaneous fission rates are determined as explained
  by~\citet{Martinez-Pinedo.Mocelj.ea:2007}.  Experimental rates for
  alpha and beta decay have been obtained from the NUDAT
  database.\footnote{\url{http://www.nndc.bnl.gov/nudat2/}} Fission
  yields for all fission processes are determined using the
  statistical code
  ABLA~\citep{Gaimard.Schmidt:1991,Benlliure.Grewe.ea:1998}.  All
heating is self-consistently added to the entropy of the fluid
following the procedure of Freiburghaus et al.~(1999).  The change of
temperature during the initial expansion is determined using the
Timmes equation of state (Timmes \& Arnett 1999), which is valid below the density $\rho \sim 3\times 10^{11}$ g cm$^{-3}$ at which our calculation begins.  

As in the $r$-process calculations performed by Freiburghaus et
al.~(1999), we use a Lagrangian density $\rho(t)$ taken from the NS-NS
merger simulations of Rosswog et al.~(1999).  In addition to $\rho(t)$, the initial temperature $T$, electron
fraction $Y_{e}$, and seed nuclei properties ($\bar{A}$,$\bar{Z}$) are
specified for a given calculation.  We assume an initial temperature $T = 6\times 10^{9}$ K, although the subsequent $r$-process heating is
not particularly sensitive to this choice because any initial thermal energy is rapidly lost to PdV work during the initial expansion before the $r$-process begins (Meyer 1989; Freiburghaus et al.~1999).  For our fiducial model we also assume $Y_{e} = 0.1$,
$\bar{Z} \simeq 36$, $\bar{A} \simeq 118$ (e.g. Freiburghaus et
al.~1999).

Our results for the total radioactive power $\dot{E}$ with time are
shown in Figure \ref{fig:edot}.  On timescales of interest the radioactive power can be divided into two contributions: fission and $\beta-$decays, which are
denoted by {\it dashed} and {\it dotted} lines, respectively.  The
large heating rate at very early times is due to the $r$-process,
which ends when neutrons are exhausted at $t \sim 1$ s $\sim 10^{-5}$
days.  The heating on longer timescales results from the synthesized
isotopes decaying back to stability.  On the timescales of interest
for powering EM emission ($t_{\rm peak} \sim$ hours--days;
eq.~[\ref{eq:tpeak}]), most of the fission results from the
spontaneous fission of nuclei with $A \sim 230-280$.  This releases
energy in the form of the kinetic energy of the daughter nuclei and fast neutrons, with a modest contribution from gamma-rays.  The other source of radioactive heating is $\beta-$decays of $r$-process product nuclei and fission
daughters (see Table \ref{table:betadecaynuclei} for examples
corresponding to our fiducial model).  In
Figure \ref{fig:edot} we also show for comparison the radioactive
power resulting from an identical mass of $^{56}$Ni and its daughter $^{56}$Co.  Note that
(coincidentally) the radioactive power of the $r$-process ejecta and $^{56}$Ni/$^{56}$Co are comparable on timescales $\sim 1$ day.

%%%%%%%%%%%%%%%%%%%%%%%%%%%%%%%%%%%%%% FIG 3 %%%%%%%%%%%%%%%%%%%%%%%%%%%%%%
\begin{figure}
\resizebox{\hsize}{!}{\includegraphics[]{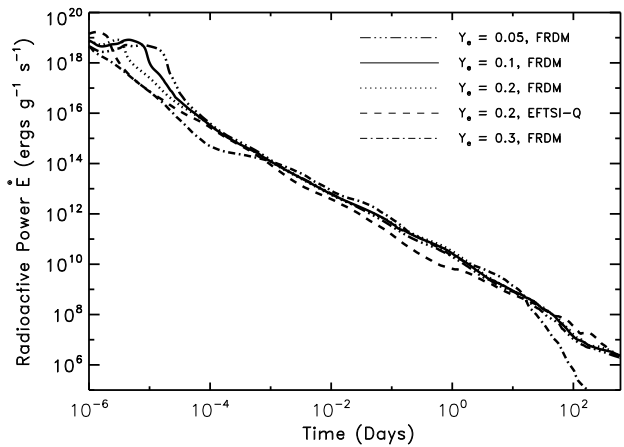}}
\caption{Total radioactive heating rate per unit mass $\dot{E}$, calculated for
  several values of the electron fraction $Y_{e}$ of the ejecta and
  for different nuclear mass models (see text).  Calculations
  employing the FRDM mass model~\citep{Moeller.Nix.ea:1995} are shown
  for $Y_{e} = 0.05$ ({\it triple dot-dashed line}), $Y_{e} = 0.1$ ({\it solid line}), $Y_{e} = 0.2$ ({\it dotted
    line}), and $Y_{e} = 0.3$ ({\it dot-dashed line}).  A calculation
  employing the EFTSI-Q~\citep{Pearson.Nayak.Goriely:1996} mass model
  is shown for $Y_{e} = 0.2$ ({\it dashed line}).  Note that on
  timescales of hours$-$days, the radioactive heating rates in
  all models agrees to within a factor $\sim 4$.}
\label{fig:edot2}
\end{figure}
%%%%%%%%%%%%%%%%%%%%%%%%%%%%%%%%%%%%%%%%%%%%%%%%%%%%%%%%%%%%%%%%%%%%%%%%%%%%
%%%%%%%%%%%%%%%%%%%%%%%%%%%%%%%%%%%%%%%%%%%%%%%%%%%%%%%%%%%%%%%%%%%%%%%%%%%%

In Figure \ref{fig:abundances} we show the final abundance distribution from our fiducial model, which shows the expected strong second and third $r$-process peaks at $A \sim 130$ and $A \sim 195$, respectively.  For comparison, we show the measured solar system $r$-process abundances with points.  The computed abundances are rather different to the one obtained by \citet{Freiburghaus+:1999} due to an improved treatment of fission yields and freeze-out effects.   
    
Although we assume $Y_{e} = 0.1$ in our fiducial model, the ejecta
from NS mergers will possess a range of electron fractions (see
$\S\ref{sec:nrichejecta}$).  To explore the sensitivity of our results
to the ejecta composition we have run identical calculations of the
radioactive heating, but varying the electron fraction in the range
$Y_{e} = 0.05-0.35$.  Although in reality portions of the ejecta with different compositions will undergo different expansion histories, in order to make a direct comparison we use the same density trajectory $\rho(t)$ as was described earlier for the $Y_{e} = 0.1$ case.  Figure \ref{fig:edot2} shows the heating rate
for ejecta with $Y_{e} = 0.05$, 0.2 and 0.3 in comparison to the
fiducial model with $Y_{e} = 0.1$.  Although the heating rate for
different values of $Y_{e}$ differs substantially at early times ($\lesssim 10^{-4}$ days), $\dot{E}$ agrees between the models to better than a factor $\sim 2$ at the later times that are the most important for transient EM emission.  

Our results for $\dot{E}$ could in principle also be sensitive to the assumed properties of the nuclei in the $r$-process
path (e.g. masses and neutron-capture cross sections), which are
uncertain and must be obtained via theoretical modeling.   
In our fiducial model (Fig.~\ref{fig:edot}) we employ the FRDM model (Moller et al.~1995) for nuclear masses.  In
order to explore the sensitivity of our results to the assumed nuclear
physics, we also performed an otherwise identical calculation using
the ETFSI-Q mass model \citep{Pearson.Nayak.Goriely:1996}, as shown in Figure
\ref{fig:edot2} for $Y_{e} = 0.2$.  Although the two models again differ in their early-time predictions for $\dot{E}$, on timescales $\gtrsim 1$ hour they converge to a heating rate within a factor $\lesssim 4$.  
%Part of the reason for this robust result is that most of the nuclei that decay on a timescale $\sim t_{\rm peak} \sim 1$ day are close to $\beta-$stability and many of their properties have been measured experimentally (see Table \ref{table:betadecaynuclei}).  

Finally, although the Lagrangian density trajectory $\rho(t)$ that we employ in our fiducial model formally corresponds to dynamically-ejected rather than wind-driven ejecta, both are likely present in NS-NS/NS-BH mergers (see $\S\ref{sec:nrichejecta}$).  Thus, we have also performed an otherwise identical calculation, but instead using a trajectory $\rho(t)$ appropriate for (higher entropy) disk winds, similar to those studied in e.g. \citet{Arcones+:2006} (cf.~\citealt*{Metzger+:2008c}; \citealt{Surman+:2008}).  Although we do not show our results explicitly in this case, we find that the heating rate $\dot{E}$ decreases in a similar manner to the dynamically-ejected material on timescales $\sim 1$ day.  However, the overall normalization of $\dot{E}$ is smaller by a factor $\sim 10$ because in high entropy winds the mass fraction of heavy nuclei is reduced at the expense of a higher alpha particle fraction, which do not contribute to the heating \citep{Hoffman+:1997}.  Although some of the wind-driven material in NS mergers may have high entropy (\citealt{Surman+:2008}; \citealt{Metzger+:2008a}), most of the total mass ejected likely has low-modest entropy ($S \sim 3-10$ k$_{\rm b}$ baryon$^{-1}$; \citealt{Metzger+:2009a}).  When correcting our results for the higher mass fraction of heavy nuclei in a lower entropy wind, we find that the heating rate $\dot{E}$ on timescales $\sim 1$ day in the wind ejecta agrees within a factor $\sim 2$ to that of the dynamically-ejected material.

To summarize, the heating rate for our fiducial model in Figure \ref{fig:edot} (which we employ throughout the remainder of the paper) appears to be relatively insensitive to the precise trajectory and composition of the ejecta, and to the uncertainties in the nuclear properties of the unstable nuclei near the $r$-process path.  

In order to understand why we find such a robust heating rate on timescales $\sim 1$ day, it is first instructive to compare $r$-process ejecta with that produced in Type Ia SNe.  In Type Ia SNe, the ejected material is processed through nuclear statistical equilibrium with $Y_{e} \approx 0.5$.  This favours the production of $N=Z$ nuclei and, in particular, $^{56}$Ni.  The $^{56}$Ni nucleus ($N=Z=28$) is produced in high abundance both because 28 is a magic nucleon number and because even-even ($N=Z$) nuclei have an additional binding energy, commonly known as the ``Wigner energy.''  At the high temperatures at which $^{56}$Ni is produced, atoms are fully ionized and, consequentially, $^{56}$Ni cannot decay by atomic electron capture.  In this case the half-life has been computed to be $t_{1/2} \approx 4\times 10^{4}$ years by ~\cite{Fisker.Martinez-Pinedo.Langanke:1999}.  Once the temperature decreases sufficiently that the inner $K$-shell orbit electrons recombine, the decay proceeds at the laboratory measured rate $t_{1/2} \simeq 6.075(10)$~days~\citep{Cruz.Chan.ea:1992}. 

The situation is different for neutron-rich $r$-process ejecta.  First, $r$-process
nuclei decay by $\beta^{-}$ and hence the half-life is unaffected by the ionization state of
the matter.  Secondly, the $r$-process results in a rather broad distribution of nuclei with 
mass number spanning the range $A \sim 110-210$.  Because the nuclei produced in NS mergers likely follow a distribution similar to their solar system abundances (see Fig.~\ref{fig:abundances}), maxima will occur at the second ($A \sim 130$) and third ($A \sim 195$) $r$-process peaks.  The {\it overall} $r$-process abundances peak in our calculations (as in the solar system) near the second peak, which is why second-peak nuclei dominate the $\beta-$decay heating rate (see Table \ref{table:betadecaynuclei}).  

We argue below, however, that the energy generation rate $\dot{E}$ is approximately {\it independent of the precise distribution of heavy nuclei}, provided that the heating is not dominated by a few decay chains and that statistical arguments can be applied.  This conclusion is supported by our results in Figure~\ref{fig:edot2}, which show that $\dot{E}$ is relatively insensitive to the composition of the ejecta, despite the fact that different electron fractions can result in rather different abundance distributions.  Perhaps most striking, the heating rate is similar whether the second $r$-process peak is produced via the fission of nuclei near the magic neutron numbers $N = 184$ ($A \sim 280$), as occurs for highly neutron-rich ejecta ($Y_{e} \lesssim 0.2$), or whether it is produced directly with little or no fission cycling, as occurs for $Y_{e} \gtrsim 0.3$.

Assuming a broad distribution of exponentially decaying nuclei with mass number $A$ the evolution of the energy generation rate $\dot{E}$ can be understood by the following arguments. For an isotopic chain of odd-A nuclei the Q-values are essentially proportional to the neutron excess $\eta \equiv N-Z$ and the beta decay rate $\lambda \propto \eta^5$ due to the 3-body
nature of the final state. The situation is slightly more complicated for even-A chains due to the presence of pairing that increases the binding energy of even-even nuclei and modifies the global proportionality of the Q-value and neutron excess. However, the selection rules of beta decay favor a maximum change in angular momentum between initial and final states of one unit; as a result, the decay of odd-odd nuclei, that typically have angular momentum $J>1$, proceeds via excited states in the daughter even-even nucleus.  Consequently, the global dependence $\lambda \propto \eta^5$ is recovered. Assuming that the number of nuclei per neutron excess interval is constant the
number of nuclei per decay rate interval, $\lambda$, and lifetime
interval, $\tau = 1/\lambda$, are given by:

\begin{equation}
  \label{eq:npertau}
  dN = N_0 \lambda^{-4/5} d\lambda, \quad dN = N_0 \tau^{-6/5} d\tau, 
\end{equation}
where $N_0$ is a normalization constant. Further assuming that the energy generation at a time $t$ is dominated by nuclei with $\tau = t$ that release energy $Q \propto \tau^{-1/5}$, the energy generation rate then becomes:

\begin{equation}
  \label{eq:edot}
  \dot{E} \propto t^{-7/5} = t^{-1.4}.
\label{eq:edotdist1}
\end{equation}

The same result can be obtained assuming that we have a distribution
of nuclei that follows equation~(\ref{eq:npertau}), each releasing an energy $Q\propto \lambda^{1/5}$ with a rate $\lambda$. In
this case the energy generation rate is:

\begin{equation}
  \label{eq:edotint}
  \dot{E} \propto \int_0^\infty \lambda \lambda^{1/5} e^{-\lambda t} N_0
  \lambda^{-4/5} d\lambda = N_0 \Gamma(7/5) t^{-7/5},
\label{eq:edotdist2}
\end{equation}
where $\Gamma$ is the Gamma function.

The above discussion neglects the fact that with increasing neutron excess the beta-decay
populates an increasing number of states in the daughter nucleus. Consequently,
we expect an exponent slightly larger than 5 for the dependence of
decay rates with neutron excess.  This will result in a power-law
decay with an exponent smaller than the value of 1.4 deduced above.  Overall, this analytic derivation is in reasonable agreement with the numerical results in Figure \ref{fig:edot}, which correspond to $\dot{E} \propto t^{-\alpha}$ with $\alpha \sim 1.1-1.3$ on timescales of hours$-$days.  Incidentally, we note that this functional form is remarkably similar to the heating rate $\dot{E} \propto t^{-1.2}$ found for the decay of nuclear waste from terrestrial reactors (\citealt{Cottingham&Greenwood:2001}; pg.~126).

\subsection{Thermalizing Processes}
\label{sec:therm}

\begin{table}
%\begin{scriptsize}
\begin{center}
\vspace{0.05 in}\caption{Properties of the dominant $\beta-$decay nuclei at $t \sim 1$ day}
\label{table:betadecaynuclei}

\begin{tabular}{lccccccc}
\hline
\hline
%{\large
\multicolumn{1}{c}{Isotope} &
\multicolumn{1}{c}{$t_{1/2}$} &
\multicolumn{1}{c}{Q$^{(a)}$} & 
\multicolumn{1}{c}{$\epsilon_{e}^{(b)}$} &
\multicolumn{1}{c}{$\epsilon_{\nu}^{(c)}$} &
\multicolumn{1}{c}{$\epsilon_{\gamma}^{(d)}$} &
\multicolumn{1}{c}{$E_{\gamma}^{\rm avg\,(e)}$} & 
\\
 & (h) & (MeV) & & & & (MeV) \\
\hline
\\
$^{135}$I & 6.57  & 2.65 & 0.18 & 0.18 & 0.64 & 1.17 \\
$^{129}$Sb & 4.4 & 2.38 & 0.22 & 0.22 & 0.55 & 0.86 \\
$^{128}$Sb & 9.0 & 4.39 & 0.14 & 0.14 & 0.73 & 0.66 \\
$^{129}$Te & 1.16 & 1.47 & 0.48 & 0.48 & 0.04 & 0.22 \\
$^{132}$I & 2.30 & 3.58 & 0.19 & 0.19 & 0.62 & 0.77 \\
$^{135}$Xe & 9.14 & 1.15 & 0.38 & 0.40 & 0.22 & 0.26 \\
$^{127}$Sn & 2.1 & 3.2 & 0.24 & 0.23 & 0.53 & 0.92 \\
$^{134}$I & 0.88 & 4.2 & 0.20 & 0.19 & 0.61 & 0.86 \\
$^{56}$Ni$^{(f)}$ & 146 & 2.14 & 0.10 & 0.10 & 0.80 & 0.53 \\

\hline
\end{tabular}
\end{center}
{\small
(a) Total energy released in the decay; (b),(c),(d) Fraction of the decay energy released in electrons, neutrinos, and $\gamma-$rays; (e) Average photon energy produced in the decay; (f) Note: $^{56}$Ni is not produced by the $r$-process and is only shown for comparison (although a small abundance of $^{56}$Ni may be produced in accretion disk outflows from NS-NS/NS-BH mergers; Metzger, Piro, $\&$ Quataert 2008). }
\end{table}

Of the total power released by nuclear reactions $\dot{E}$ (Figure \ref{fig:edot}), only a fraction $\epsilon_{\rm therm}$ will thermalize with the plasma and hence be useful for powering EM emission.  In this section we estimate $\epsilon_{\rm therm}$.  Since the light curve peaks on approximately the timescale $t_{\rm peak}$ (eq.~[\ref{eq:tpeak}]), we shall normalize our considerations to this time.

\subsubsection{$\beta$-decay Heating}
\label{sec:betadecays}

First, consider the energy released by $\beta-$decays, which dominate $\dot{E}$ at late times (Fig.~\ref{fig:edot}).  The total energy released in the decay $Q$ is divided between the outgoing neutrino and electron, and the gamma-rays produced as the daughter nucleus cascades to the ground state from excited nuclear levels.  In Table \ref{table:betadecaynuclei} we list the properties of a sample of nuclei which contribute appreciably to $\dot{E}$ at $t \sim t_{\rm peak} \sim 1 $ day, as determined from our network calculations in $\S\ref{sec:network}$.  The properties listed include the decay half-life $t_{1/2}$, the relative fraction of $Q$ carried away by the electron, neutrino, and gamma-rays ($\epsilon_{e},\epsilon_{\nu},$ and $\epsilon_{\gamma}$, respectively) and the mean gamma-ray energy $E_{\gamma}^{\rm avg}$.  Most of this information was obtained or calculated using data from the Lawrence Berkeley National Laboratory's Isotopes Project.\footnote{\url{http://ie.lbl.gov/education/}.}  

Because the energy imparted to the outgoing electron $E_{e} =
\epsilon_{e}Q$ is generally greater than or similar to the electron
rest mass (0.511 MeV), the electron is mildly relativistic and, as a
result, carries a similar fraction of the outgoing energy as the
neutrino (i.e. $\epsilon_{e} \approx \epsilon_{\nu}$).  Although the
neutrino readily escapes the ejecta and does not contribute to the
heating, the electron is charged and interacts electromagnetically
with the ambient electrons and nuclei.  The dominant thermalizing
process is electron-electron coulomb scattering.

For electron-electron scattering in the fast test particle limit
($E_{e} \gg kT$, where $T \sim 10^{4}$ K is the temperature of the
background plasma), the energy exchange (or ``thermalization'')
timescale is given by \be t_{\rm therm}^{e-e} \approx 4.6\times
10^{13}{\,\rm s}\left(\frac{E_{e}}{0.5{\,\rm MeV}}\right)^{3/2}{\rm
  ln}\Lambda^{-1}n_{e}^{-1}, \ee where ln$\Lambda$ is the Coulomb
logarithm (e.g. NRL Plasma Formularly; \citealt{Huba:2007}).  Assuming a
spherical, homogeneous outflow, the electron density $n_{e}$ at time
$t$ is approximately given by 
\begin{eqnarray} && n_{e} = \frac{M_{\rm ej}}{(4\pi/3)R^{3}\mu_{e}} \nonumber \\
&& \approx 
10^{12}{\rm\,cm^{-3}}\left(\frac{M_{\rm
      ej}}{10^{-2}M_{\sun}}\right)^{-1/2}\left(\frac{v}{0.1
    c}\right)^{-3/2}\left(\frac{t}{t_{\rm peak}}\right)^{-3},
\label{eq:ne}
\end{eqnarray}
where $\mu_{e} \approx Am_{n}/Z$ is the mean mass per electron, $m_{n}$ is the mass of a nucleon, and we have assumed an average charge $Z \sim 60$ and mass $A \sim 130$ for the $r$-process nuclei.  Although the $r$-process nuclei are only partially ionized on timescales $t \sim t_{\rm peak}$, $n_{e}$ includes both free and bound electrons because, for purposes of high energy scattering, they have identical cross sections (the impact parameter for $E_{e} \sim $ MeV is much smaller than the atomic scale).

Thus, the ratio of the thermalization time due to electron-electron collisions  $t_{\rm th}^{e-e}$ to the timescale at which the emission peaks is given by
\begin{eqnarray}
&&t_{\rm therm}^{e-e}/t_{\rm peak} \approx \nonumber\\
 &&10^{-4}\left(\frac{E_{e}}{\rm 0.5\,MeV}\right)^{3/2}\left(\frac{\rm{ln}\Lambda}{10}\right)^{-1}\left(\frac{v}{0.1\, c}\right)^{2}\left(\frac{t}{t_{\rm peak}}\right)^{3}.
\label{eq:t_therm_ee}
\end{eqnarray}
Equation (\ref{eq:t_therm_ee}) shows that for a typical value $E_{e} \sim$ 0.5 MeV, $t_{\rm therm}^{e-e} > t_{\rm peak}$ for $t \lesssim 10$ $t_{\rm peak}$, implying that the $\beta-$decay electrons will efficiently thermalize on the timescales of interest.  
%Note: I calculate Log_Lambda = 16

Table $\ref{table:betadecaynuclei}$ shows that typically $\sim 1/2$ of
the $\beta-$decay energy is also released in the form of $\sim$ MeV
gamma-rays.  Although a portion of the gamma-rays will thermalize via
Compton scattering and photoelectric absorption (e.g. \citealt{Colgate+:1980}; \citealt{Swartz+:1995}), a significant fraction will also
escape, especially as the ejecta expands and the optical depth
decreases.  In the case of $^{56}$Ni and $^{56}$Co, for example,
\citet{Swartz+:1995} find an effective absorptive opacity which is
about $\sim 15$ per cent of the fully-ionized Thomson opacity
(i.e. $\kappa_{\gamma} \approx 0.03$ cm$^{2}$g$^{-1}$) using Monte
Carlo radiative transfer calculations.  Since $t_{\rm peak}$ is
attained when the optical depth is $\tau_{\rm peak} \sim c/v \sim 10$,
the thermalization ``optical depth'' is $\lesssim 1$ for times greater
than $\approx (\kappa/\kappa_{\gamma})(v/c)t_{\rm peak} \sim t_{\rm
  peak}$ (see eqs.~[\ref{eq:tdiff}] and [\ref{eq:tpeak}]).  As Table
$\ref{table:betadecaynuclei}$ illustrates, the mean $\gamma-$ray
energy $E_{\gamma}^{\rm avg}$ from $^{56}$Ni decay is within a factor
$\sim 2$ of those produced by the other $\beta-$decays, so we expect
similar $\gamma-$ray thermalization properties in Type Ia SNe and in
NS merger ejecta.  We conclude that photons will partially thermalize
for $t \lesssim t_{\rm peak}$, but they will contribute little heating
at later times ($t \gg t_{\rm peak}$).

From Table $\ref{table:betadecaynuclei}$ we infer average values of $\epsilon_{e} \approx \epsilon_{\nu} \approx 0.25$ and $\epsilon_{\gamma} \approx 0.5$ (Table \ref{table:betadecaynuclei}).  Combining our results, we conclude that the effective $\beta-$decay thermalization fraction will vary from $\epsilon_{\rm therm} \approx \epsilon_{e}+\epsilon_{\gamma} \approx 0.75$ to $\epsilon_{\rm therm} \approx \epsilon_{e} \approx 0.25$ as the ejecta expands from $R \ll R_{\rm peak}$ to $R \gtrsim R_{\rm peak}$.

\subsubsection{Fission Heating}
\label{sec:fission}

In the case of fission most of the radioactive energy is released as kinetic energy of the fission product nuclei, with a typical daughter energy of $E_{A} \sim 100$ MeV.  In this case the dominant thermalizing process is coulomb scattering off ambient nuclei of similar mass A and charge Z.  In the case of ion-ion collisions the thermlization timescale is \citep{Huba:2007}
\begin{eqnarray}
&t_{\rm therm}^{A-A}& \approx \nonumber \\
&5\times 10^{12}{\,\rm s}&{\rm ln}\Lambda^{-1}n_{A}^{-1}\left(\frac{E_{A}}{100{\,\rm MeV}}\right)^{3/2}\left(\frac{A}{130}\right)^{1/2}
 \left(\frac{Z}{60}\right)^{-4},
\end{eqnarray}
where $n_{A} \approx \rho/Am_{n}$ is the number density of ambient nuclei.  Thus, using equations (\ref{eq:tpeak}) and (\ref{eq:ne}) the ratio of the thermalization timescale to the timescale at which the emission peaks is given by
\begin{eqnarray}
t_{\rm therm}^{A-A}/t_{\rm peak} \approx  6\times 10^{-3}\left(\frac{E_{A}}{\rm\,\, 100 MeV}\right)^{3/2}\left(\frac{A}{130}\right)^{3/2}\times \nonumber \\
 \left(\frac{Z}{60}\right)^{-4}\left(\frac{{\rm ln}\Lambda}{10}\right)^{-1}\left(\frac{v}{0.1 c}\right)^{2}\left(\frac{t}{t_{\rm peak}}\right)^{3}. 
\label{eq:t_therm_AA}
\end{eqnarray}
Since $t_{\rm therm}^{A-A} \ll t_{\rm peak}$ we conclude that the fission daughters will also thermalize on timescales $\sim t_{\rm peak}$, implying that $\epsilon_{\rm therm} \approx 1$ for fission.  Therefore, even though fission contributes less to $\dot{E}$ than $\beta-$decays at $t \sim t_{\rm peak} \sim 1$ day, its higher thermalized fraction suggests that it may dominate the heating.

\subsubsection{Neutron Heating}

Neutrons, emitted either spontaneously following fission or induced by $\beta-$decays, also carry a modest portion of the released energy.  Although the contribution of neutrons to $\dot{E}$ is generally much less than that of $\beta-$decays or the kinetic energy of fission daughters, we consider their thermalization as well for completeness.

At high densities (e.g. in terrestrial reactors) fission-product
neutrons can be captured by heavy nuclei to induce further reactions.
For the much lower densities of present interest, however, the neutron
capture timescale (with a typical cross section $\sim$ millibarns) is
much longer than the free neutron beta-decay timescale $\sim
15$ minutes.  As a result, fast neutrons created by fission rapidly
decay into into fast protons (with a typical energy $E_{p} \sim 1$
MeV) before capturing.  In order to thermalize, the protons must
exchange energy with the much heavier ambient charged particles (the
proton density is much too low for p-p scattering to be efficient).
We find that the proton's thermalization time is larger by a factor
$\sim Z^{2}A^{1/2}(E_{p}/E_{A})^{3/2} \sim 10^{2}$ than that of the
fission daughter nuclei (eq.~[\ref{eq:t_therm_AA}]).  This suggests
that the proton will have $t_{\rm therm} \sim t_{\rm peak}$ and hence
may not thermalize.
%GMP should one also not consider the proton-electron interactions for
%thermalization. I read somewhere that under Big Bang conditions this
%is what keeps protons in thermal equilibrium. 

Our estimates above neglect, however, the possible effects of magnetic
fields, which can trap charged particles and enhance their
thermalization if the field is directed perpendicular to the outflow
velocity (e.g. \citealt{Colgate+:1980}).  For instance, if the NS involved
in the merger has a (modest) surface field strength of $B \approx
10^{9}$ G, this translates into a field strength of $B \sim 3 \times
10^{-8}$ G at $R \sim R_{\rm peak}$ by flux freezing.  The larmor
radius for a 1 MeV proton at $R_{\rm peak}$ is $r_{\rm L} \sim
10^{12}$ cm, which is $\ll R_{\rm peak} \sim 10^{14}$ cm.  This
suggests that the proton's residence time (and hence thermalization)
may be significantly enhanced due to the magnetic field.  As a result,
we conclude that the energy released in neutrons will also likely
thermalize.

\subsubsection{Net Heating Efficiency and the Effective Value of f}
\label{sec:fcalc}

Considering both $\beta-$decays (with $\epsilon_{\rm therm} \approx 0.25-0.75$) and fission ($\epsilon_{\rm therm} \approx 1$), we conclude that the net heating fraction of merger ejecta is between $\epsilon_{\rm therm} \approx 0.25$ and $\approx 1$, depending on time and the relative contributions of $\beta-$decays and fission to $\dot{E}$. 

From Figure \ref{fig:edot} we find that $\dot{E}$ decreases approximately as a power law $\dot{E} \propto t^{-\alpha}$, with a value $\alpha \sim 1.1-1.4$ on timescales of hours$-$days, relatively close to the heating functional form adopted by LP98: $\dot{Q} \propto fc^{2}/t$ (see eq.~[\ref{eq:lpeak}] and surrounding discussion).  In $\S\ref{sec:network}$ we presented a simple derivation of this result that explains why $\alpha \lesssim 1.4$ (see eqs.~[\ref{eq:edotdist1}]-[\ref{eq:edotdist2}]).  For an average thermalization efficiency of $\epsilon_{\rm therm} \sim 0.75$, our results imply that the {\it effective value} of $f$ is $\approx 3\times 10^{-6}$ at $t \sim 1$ day.  This is somewhat lower than the range of values considered by LP98 and much lower than has been estimated elsewhere in the literature.  For instance, \citet{Rosswog:2005} estimates a NS merger transient peak luminosity $L_{\rm peak} \sim 10^{44}$ ergs s$^{-1}$, corresponding to an effective value of $f \sim 10^{-3}$ for $M_{\rm ej} \sim 10^{-2}M_{\sun}$ (eq.~[\ref{eq:lpeak}]).  He derives this under the assumption that an appreciable fraction of the total energy released in forming heavy $r$-process nuclei ($B/A \sim 8$ MeV nucleon$^{-1}$) is released over a timescale $t_{\rm peak} \sim 1$ day.  This is incorrect because {\it most} of the binding energy is released by the formation of seed nuclei in the initial expansion (on timescales $\sim$ milliseconds) and by the subsequent $r$-process (on timescales $\lesssim 1$ second; see Fig.~\ref{fig:edot}).  Since this heating all occurs at radii $R \lesssim (v\times 1 {\rm\,second}) \sim (v/0.3 c)10^{10}$ cm, this early-time heating suffers a factor $\gtrsim 10^{4}$ loss in thermal energy due to PdV work before the outflow expands to the radius $R_{\rm peak} \sim 10^{14}$ cm (eq.~[\ref{eq:rpeak}]) at which photons can finally escape.  Instead, the luminosity at $t \sim t_{\rm peak}$ is primarily powered by {\it residual} energy released as $r$-process products fission and decay back to stability.  This occurs on much longer timescales and involves a significantly smaller energy release (typically closer to $\sim 10^{-3}$ MeV nucleon$^{-1}$ on timescales $\sim 1$ day).  

%\begin{eqnarray}
%&t_{\rm therm}^{A-p} &\approx \frac{2^{3/2}m_{p}^{1/2}E_{p}^{3/2}}{8\pi n Z^{4}e^{4}ln\Lambda} \approx\nonumber \\

%\end{eqnarray}

\section{Light Curves}
\label{sec:lightcurves}

\subsection{Radiative Transfer Calculation}
\label{sec:sedona}

We calculate the light curves of compact object merger ejecta using the time-dependent radiative transfer code SEDONA \citep{Kasen+:2006}.  Our set-up is similar to that described in Darbha et al.~(2010, in prep) for the case of $^{56}$Ni decay-powered transients produced by the accretion-induced collapse of white dwarfs.  However, in the present case we have modified the code to include the radioactive heating $\dot{Q} = \epsilon_{\rm therm}\dot{E}$ for NS mergers as calculated in $\S\ref{sec:heating}$.  Although SEDONA can track $\gamma-$ray thermalization, we do not use this option given the large number of decaying nuclei and their complex $\gamma$-ray spectra; rather, we incorporate the escape fraction into an average thermalization efficiency $\epsilon_{\rm therm}$ that we hold constant in time.  We calculate models assuming both $\epsilon_{\rm therm} = 0.5$ and $\epsilon_{\rm therm} =1$ in order to bracket the uncertainty in the precise fraction of $\gamma-$rays that thermalize (see $\S\ref{sec:fcalc}$). 

Although the ejecta from NS-NS/NS-BH mergers is likely to be highly asymmetric (e.g. the ``banana-like'' geometry of tidal tails; Rosswog 2005), we assume a one-dimensional (spherically symmetric) geometry for simplicity.  We shall address the effects of the full multi-dimensional kinematics of the outflow in future work.  Since SEDONA uses a velocity-time grid, a required input to the calculation is the velocity profile $\rho(v)$ of the (assumed homologous) expansion.  We take the density profile to decrease as $\rho \propto v^{-3}$ between $v \approx 0.05-0.2$ c, as motivated by the typical velocity $v \sim 0.1$ c of the dynamically-ejected and wind-driven neutron-rich ejecta from NS-NS/NS-BH mergers ($\S\ref{sec:nrichejecta}$).  We assume that $\rho(v)$ decreases exponentially with $v$ outside this range and have verified that our results are not sensitive to precisely how we taper the edges of the velocity distribution.  We find that our results are also similar if we instead assume $\rho \propto v^{-2}$ or $\rho \propto v^{-4}$, implying that the exact density profile is not crucial for the overall light curve shape.

Another input to our calculation is the composition, the ionization
energies of each element, and the bound-bound and bound-free opacities
of each element.  Unfortunately, the spectral line information for
these very high-Z elements is very limited.  Most of the data
available is experimental \citep[e.g.][]{Lawler+:2006,Lawler+:2007,Lawler+:2009,Biemont+:2007}, since many body quantum mechanical
calculations of these elements' spectra represent a formidable task,
even with modern computing.  Much of the experimental work has focused
on aiding studies of $r$-process abundances in ultra metal-poor halo
stars, which generally make use of resonant absorption lines at
optical wavelengths (e.g. \citealt{Cayrel:1996}; \citealt{Sneden+:2003}).  However,
the total {\it opacity} of most relevance to merger transients results
from densely-packed UV lines, for which there is currently
insufficient information in either the Kurucz line list \citep{Kurucz:1995}
or the more recent experimental studies.  Nevertheless, the spectra of
at least some high-Z $r$-process elements are almost certainly as
complex as Fe peak elements, if not more (G.~Wahlgren, private
communication); this is important because the Fe peak elements cause
UV ``line-blanketing'' in normal SN spectra.  We expect that the same
effect is likely to be produced by third $r$-process peak elements
since they are largely transition metals.

Given this lack of spectral information, we attempt to crudely account for the effects of the unknown $r$-process element lines on the opacity by using the bound-bound lines of Fe, but modified to include the correct ionization energies of the $r$-process elements.  Specifically, our calculations use the ionization energies of Pb as a representative $r$-process element.   These uncertainties in the bound-bound transitions obviously limit our ability to make detailed spectroscopic predictions, but it does allow us to qualitatively address the effects of line blanketing on the transients' color evolution.  In addition, the overall lightcurve shape, the peak luminosity, and the characteristic timescale of the event ($\sim$ day) are robust in spite of these uncertainties.

\subsection{Results}
\label{sec:results}

%%%%%%%%%%%%%%%%%%%%%%%%%%%%%%%%%%%%%% FIG 3 %%%%%%%%%%%%%%%%%%%%%%%%%%%%%%
\begin{figure}
\resizebox{\hsize}{!}{\includegraphics[]{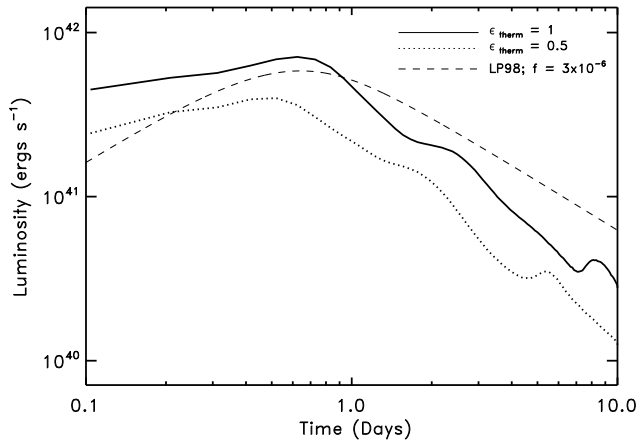}}
%\resizebox{\hsize}{5cm}{\includegraphics{1a.eps}}
%\includegraphics{1a.eps}
\caption{Bolometric light curve of the radioactively-powered transients from NS-NS/NS-BH mergers, calculated assuming a total ejecta mass $M_{\rm tot} = 10^{-2}M_{\sun}$ with electron fraction $Y_{e} = 0.1$ and mean outflow speed $v \simeq 0.1$ c, and for two values of the thermalization efficiency ($\S\ref{sec:therm}$), $\epsilon_{\rm therm} = 1$ ({\it solid line}) and $\epsilon_{\rm therm} = 0.5$ ({\it dotted line}).  Also shown for comparison ({\it dashed line}) is a one-zone calculation based on the LP98 model (as implemented in \citealt{Kulkarni:2005} and \citealt*{Metzger+:2008c}) assuming $f = 3\times 10^{-6}$ (see $\S\ref{sec:fcalc}$) and the same values for $M_{\rm tot}$ and $v$. }
\label{fig:lc}
\end{figure}
%%%%%%%%%%%%%%%%%%%%%%%%%%%%%%%%%%%%%%%%%%%%%%%%%%%%%%%%%%%%%%%%%%%%%%%%%%%%
%%%%%%%%%%%%%%%%%%%%%%%%%%%%%%%%%%%%%%%%%%%%%%%%%%%%%%%%%%%%%%%%%%%%%%%%%%%%

%%%%%%%%%%%%%%%%%%%%%%%%%%%%%%%%%%%%%% FIG 3 %%%%%%%%%%%%%%%%%%%%%%%%%%%%%%
\begin{figure}
\resizebox{\hsize}{!}{\includegraphics[]{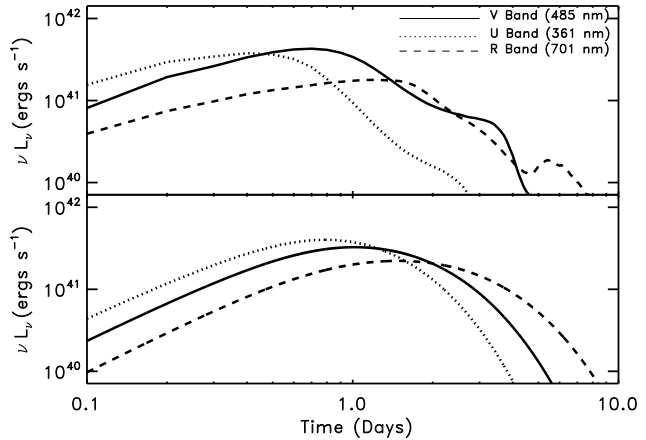}}
%\resizebox{\hsize}{5cm}{\includegraphics{1a.eps}}
%\includegraphics{1a.eps}
%GMP: Maybe is the standard notation in this community but I find
%confusing the use of \nu L_\nu. It is not enough to denote as L_\nu
%the luminosity in a particular wavelength. I will not expect \nu
%L_\nu to have the same units than L. 
\caption{{\it Top Panel}: $\nu L_{\nu}$ color light curves from the
  $\epsilon_{\rm therm} = 1$ calculation in Figure \ref{fig:lc}.
  V$-$, U$-$, and R$-$ band light curves are shown with solid, dotted,
  and dashed lines, respectively.  {\it Bottom Panel}: Analogous color
  evolution predicted by the LP98 blackbody model.}
\label{fig:colors}
\end{figure}
%%%%%%%%%%%%%%%%%%%%%%%%%%%%%%%%%%%%%%%%%%%%%%%%%%%%%%%%%%%%%%%%%%%%%%%%%%%%
%%%%%%%%%%%%%%%%%%%%%%%%%%%%%%%%%%%%%%%%%%%%%%%%%%%%%%%%%%%%%%%%%%%%%%%%%%%%

Figure \ref{fig:lc} shows our results for the bolometric light curve for a fiducial model with $Y_{e} = 0.1$, ejecta mass $M_{\rm ej} = 10^{-2}M_{\sun}$, and outflow speed $v = 0.1$ c.  We show two calculations performed using different values for the assumed thermalization efficiency, $\epsilon_{\rm therm} = 0.5$ and $\epsilon_{\rm therm} = 1$, which roughly bracket our uncertainty in the $\gamma-$ray escape fraction ($\S\ref{sec:therm}$).  Also shown for comparison in Figure \ref{fig:lc} is the toy model of LP98 (cf.~\citealt{Kulkarni:2005}; $\S\ref{sec:LPanalytics}$), calculated assuming an electron scattering opacity and an ``f'' value $= 3\times 10^{-6}$, as calibrated to match the radioactive heating rate in $\S\ref{sec:fcalc}$.  Figure \ref{fig:lc} shows that the light curves predicted using the toy model and our more detailed calculation are relatively similar near the time of peak emission ($t_{\rm peak} \sim 1$ day), but their differences become more pronounced at earlier and later times.  The ``bumps'' in the light curve at $t \sim $ few days in our calculation are due to recombination of the outer shell electrons in our representative high-Z element Pb (and the resulting opacity change) as the expanding photosphere cools.  

In the top panel of Figure $\ref{fig:colors}$ we decompose the light curve into luminosities $\nu L_{\nu}$ in several standard optical bands (i.e. ``colors'').  The bottom panel of Figure $\ref{fig:colors}$ shows the analogous color evolution predicted with the LP98 model, which assumes a perfect (single temperature) blackbody spectrum.  Both calculations predict that the light curve peaks earlier in time at shorter wavelengths because the photospheric temperature decreases with time as the ejecta expands.  However, the LP98 model predicts an overall $\nu L_{\nu}$ peak in the UV, while our calculation predicts an earlier peak at longer wavelengths (i.e.~in V-band) and a clear suppression in UV emission at times $t \gtrsim 1$ day.  This behavior results from strong UV absorption due to dense bound-bound transitions (``line blanketing''), which produces a much redder spectrum than would be predicted by assuming a grey opacity.  Indeed, rapid reddening following the peak emission epoch is likely a defining characteristic of kilonovae.

We have also explored the sensitivity of our results to the mass of the ejecta by performing an otherwise identical calculation, but with a lower ejecta mass $M_{\rm ej} = 10^{-3}M_{\sun}$.  Our results for the color evolution are shown in Figure \ref{fig:colors2}.  Although the qualitative features of the light curve evolution are similar to the $M_{\rm ej} = 10^{-2}M_{\sun}$ case, the V band light curve peaks somewhat earlier and at a lower luminosity, as expected from the analytic scaling relationships $t_{\rm peak} \propto M_{\rm ej}^{1/2}$ (eq.~[\ref{eq:tpeak}]) and $L_{\rm peak} \propto M_{\rm ej}^{1/2}$ (eq.~[\ref{eq:lpeak}]).  The higher photosphere temperature at the epoch of peak emission for lower $M_{\rm ej}$ ($T_{\rm peak} \propto M_{\rm ej}^{-1/8}$; eq.~[\ref{eq:temppeak}]) also results results in somewhat bluer peak emission.

\section{Detection Prospects}
\label{sec:detection}

Because the radioactively-powered emission from NS-NS/NS-BH mergers is relatively isotropic, it can in principle be detected in at least three independent ways: (1) coincident with a source of detected GWs; (2) coincident with a short-duration GRB, and (3) via blind transient surveys (e.g. PTF and LSST).  In this section, we discuss each possibility in turn.

\subsection{Gravitational Wave-Triggered Follow-Up}

Advanced LIGO is expected to be sensitive to NS-NS and NS-BH mergers out to a distance $\sim 300$ and $\sim 650$ Mpc, respectively \citep{Cutler.Thorne:2002}.  For an ejecta mass of $M_{\rm ej} = 10^{-2}M_{\sun}$ we predict a peak V-band luminosity of $\approx 3\times 10^{41}$ ergs s$^{-1}$ (Figure \ref{fig:lc}), corresponding to an absolute magnitude $M_{V} \sim -15$.  Thus, the entire Advanced LIGO volume could be probed by searching down to magnitude $V \sim 22-24$.  Although this represents a realistic depth for a moderately large telescope, the positional uncertainty of LIGO/Virgo detections is expected to range from many arcminutes to degrees (e.g. \citealt{Sylvestre:2003}).  As a result, both sensitivity and a large field of view (i.e. a high ``\'etendue'') are requirements for any follow-up instrument.  Since these figures of merit are already optimized for transient survey telescopes, projects such as PTF and (eventually) LSST and SASIR may be optimal for GW follow-up (in addition to their role in blind transient searches; $\S\ref{sec:surveysearch}$).

%%%%%%%%%%%%%%%%%%%%%%%%%%%%%%%%%%%%%% FIG 3 %%%%%%%%%%%%%%%%%%%%%%%%%%%%%%
\begin{figure}
\resizebox{\hsize}{!}{\includegraphics[]{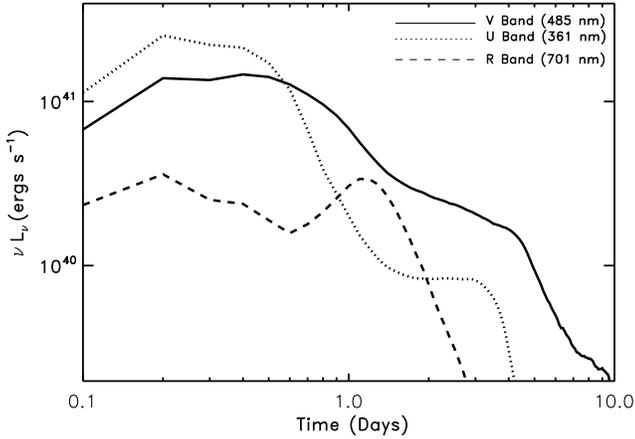}}
\caption{Same as the top panel in Figure \ref{fig:colors}, but calculated for $M_{\rm ej} = 10^{-3}M_{\sun}$.}
\label{fig:colors2}
\end{figure}
%%%%%%%%%%%%%%%%%%%%%%%%%%%%%%%%%%%%%%%%%%%%%%%%%%%%%%%%%%%%%%%%%%%%%%%%%%%%
%%%%%%%%%%%%%%%%%%%%%%%%%%%%%%%%%%%%%%%%%%%%%%%%%%%%%%%%%%%%%%%%%%%%%%%%%%%%

Given the short-lived duration $\sim 1$ day of the expected transient signal, rapid GW data reduction (e.g. \citealt{Marka+:2002}) and dissemination of candidate detections to the astronomical community will be crucial for detection and follow-up (as has been adopted by the neutrino community; e.g. \citealt{Kowalski.Mohr:2007}; \citealt{Abbasi+:2009}; \citealt{Stamatikos+:2009}).  Indeed, given the unique optical signature of NS-NS/NS-BH merger transients (\S\ref{sec:lightcurves}), moderately deep optical/NIR follow-up of even low-significance potential GW sources could improve the effective sensitivity of Advanced LIGO/Virgo with only a relatively moderate investment of resources (see \citealt{Kowalski.Mohr:2007} for an example applied to high energy neutrino point sources).  This could prove to be particularly important if the merger rates are at the low end of current estimates.  Recently, efforts have begun to set up a rapid GW data analysis pipeline to produce location-probability sky maps within $\sim 5-10$ minutes following a GW detection with LIGO/Virgo (\citealt{Kanner+:2008}).  In fact, a pilot program for the prompt follow-up of GW triggers with wide-field optical telescopes such as QUEST and TAROT is already underway (P.~Shawhan, private communication).  

\subsection{Short-Duration GRB Follow-Up}

Given the possible association between short-duration GRBs and NS-NS/NS-BH mergers, another method to search for kilonovae is with deep optical/IR observations on $\sim 1$ day timescale following the burst.  Recently, \citet{Kocevksi+:2009} presented upper limits on the presence of a putative LP98 kilonova using follow-up observations of GRB 070724A and GRB 050509b \citep{Hjorth+:2005}.  Our results in $\S\ref{sec:results}$ show that the LP98 model (properly calibrated) does a reasonably good job of reproducing the qualitative features of the optical/NIR light curves around the time of peak emission.  Thus, by assuming $v \approx 0.1$ c and $f \approx 3\times 10^{-6}$ ($\S\ref{sec:fcalc}$) we conclude from their Figures 8 and 9 that $M_{\rm ej}$ must be $\lesssim 0.1 M_{\sun}$ and $\lesssim 10^{-3}M_{\sun}$ for GRBs 070724A and 050509b, respectively.  Although the former (070724A) is not particularly constraining on merger models because such a large ejecta mass is unlikely, the latter non-detection (050509b) suggests that the disk that formed in this event was rather small ($\lesssim 10^{-2}M_{\sun}$; see the discussion in $\S\ref{sec:nrichejecta}$).  A low disk mass is, however, still consistent with a merger interpretation for this event because the isotropic luminosity of the GRB was quite low ($E_{\gamma,\rm iso} \sim 2.4\times 10^{48}$ ergs; \citealt{Kann+:2008}): even ignoring geometric beaming corrections, only an accreted mass $\sim 10^{-5}M_{\sun}$ is necessary to produce a relativistic jet with energy $E_{\gamma,\rm iso}$ assuming that the efficiency for converting rest-mass energy to $\gamma-$ray power is $\sim 10$ per cent (e.g. \citealt{McKinney:2005}).

\citet{Berger+:2009} present additional early optical/NIR follow-up observations of GRB 070724a, including the discovery of transient emission peaking $t \sim 3$ hours following the burst.  Due to the transient's very red spectrum (which is highly unusual for a standard GRB afterglow) they suggest a possible NS merger transient interpretation.  They conclude, however, that this possibility is unlikely: the brightness and rapid rise time of the transient require values of $f \sim 5\times 10^{-3}$ and $M_{\rm ej} \sim 10^{-4}M_{\sun}$ which, within the standard LP98 model, predicts a photospheric temperature peaked in the UV (eq.~[\ref{eq:tpeak}]) and thus inconsistent with the transient's red colors.  Our calculations in $\S\ref{sec:lightcurves}$ suggest that UV line blanketing could produce a redder spectrum (see Fig.~\ref{fig:colors}).  However, the value of $f \sim 5\times 10^{-3}$ they nominally require is three orders of magnitude higher than we predict from radioactive decay ($\S\ref{sec:fcalc}$).  It thus appears more likely that the emission detected by Berger et al.~is afterglow-related (with the observed reddening perhaps due to dust).  

Perhaps the most promising candidate kilonova detection to date was following GRB 080503, which showed an unusual rise in its optical afterglow light curve at $t \sim 1$ day before rapidly fading over the next several days \citep{Perley+:2009}.  Although limited color information was obtained near the emission peak, the observed light curve evolution is largely consistent with that expected from a kilonova for an assumed ejecta velocity $v \sim 0.1$ c and mass $M_{\rm ej} \sim$ few $\times 10^{-2}M_{\sun}$.  However, although the event was well-localized on the sky, no obvious host galaxy was detected coincident with the burst, despite the fact that a relatively low redshift $z \sim 0.1$ would be required to fit the observed peak brightness (if the event was indeed a kilonova).  In principle, NS-NS/NS-BH binaries can receive natal ``kicks'' from their supernovae which may eject them into intergalactic space, thereby making it difficult to identify their original host galaxy.  This possibility is consistent with the very low density of the circumburst medium inferred for 080503.  Perhaps more problematic for the kilonova interpretation, however, is the X-ray detection by {\it Chandra} coincident with the optical rise, which suggests that the optical emission at $\sim 1$ day may simply be due to an (albeit unusual) non-thermal afterglow.  The fact that the X-ray luminosity exceeds the optical luminosity appears especially difficult to explain if both are related to radioactively powered quasi-thermal emission.

Although the possible presence of kilonovae following short GRBs is not well-constrained at present, this situation could in principle improve with additional sensitive, early-time optical/NIR observations of short GRBs.  Indeed, {\it Swift} should remain operational through the next decade and presently detects short GRBs at a rate of about $\sim 10$ per year.  Unfortunately, however, this approach may encounter fundamental obstacles due to the non-thermal afterglow emission which also generally accompanies GRBs.  In most accretion-powered GRB models, the luminosity of the burst increases with the accretion rate (e.g. \citealt{McKinney:2005}; \citealt{Zalamea.Beloborodov:2009}) and, hence, with the disk mass.  Since the quantity of neutron-rich ejecta may be a relatively constant fraction of the disk mass (\citealt{Metzger+:2009a}; $\S\ref{sec:nrichejecta}$), the luminosity of the kilonova ($L_{\rm peak} \propto M_{\rm ej}^{1/2}$; eq.~[\ref{eq:lpeak}]) may positively correlate with the luminosity of the GRB.\footnote{An exception may occur in the case of BH-NS mergers, where in some cases large amounts of material can be ejected relative to the mass of the accretion disk that forms \citep{Rosswog:2005}}  Since the afterglow luminosities of short GRBs appears to scale with the prompt GRB fluence (as in long-duration GRBs; \citealt{Nysewander+:2009}), the afterglow may generically swamp any putative kilonova emission.  To date, this appears to be true even in cases when the circumburst density appears to be very low and the afterglow is relatively dim \citep{Perley+:2009}.

\subsection{Blind Optical Transient Surveys}
\label{sec:surveysearch}

In this section we assess the prospects for detecting kilonovae from NS-NS/NS-BH mergers with present and upcoming optical transient surveys.  The virtue of such a search strategy is that it does not rely on a GW or high-energy EM trigger.

Based on observed binary NS systems, \citet{Kalogera+:2004} find that the NS-NS merger rate in the Milky Way is between $1.7\times 10^{-5}$ and $2.9\times 10^{-4}$ yr$^{-1}$ at 95$\%$ confidence.  Population synthesis estimates (e.g., \citealt{Belczynski+:2006}) are consistent with this range but with larger uncertainties.  Since there are no known BH-NS binaries, the BH-NS merger rate is even less certain.  \citet{Bethe.Brown:1998} argue that BH-NS mergers could be substantially more common than NS-NS mergers, with \citet{Bethe+:2007} estimating a rate $\sim 10^{4}$ Gpc$^{-3}$yr$^{-1}$, corresponding to $\sim 10^{-3}$ yr$^{-1}$ in the Milky Way.  An interesting limit can be placed on the total amount of neutron-rich ejecta from NS-NS/NS-BH merger from Galactic chemical evolution (e.g. \citealt{Metzger+:2009a}).  Accounting for the total observed abundances of elements with $A \gtrsim 100$ in our Galaxy, for example, requires an average production rate $\sim 10^{-6} M_{\sun}$ yr$^{-1}$ (e.g. \citealt{Qian:2000}).  Assuming that a merger ejects $M_{\rm ej} \sim 10^{-2}(10^{-3})M_{\sun}$ on average, the Galactic merger rate cannot exceed $\sim 10^{-4}(10^{-3})$ yr$^{-1}$ in order to avoid over-producing these rare neutron-rich isotopes (see further discussion in $\S\ref{sec:rprocess}$ below).

Assuming that the NS-NS rate is proportional to the blue stellar luminosity \citep{Phinney:1991,Kopparapu+:2008}, a Galactic rate of $R_{\rm NS-NS} \equiv 10^{-4}R_{-4}$ yr$^{-1}$ corresponds to a volumetric rate of $10^{-6}R_{-4}$ Mpc$^{-3}$ yr$^{-1}$.  For $v = 0.1$ c and $M_{\rm ej} \sim 10^{-2}M_{\rm ej,-2}M_{\sun}$ our calculations predict an optical transient with a peak V-band luminosity $\nu L_{\nu} \sim 3\times 10^{41}M_{\rm ej,-2}^{1/2}$ ergs s$^{-1}$ (Fig.~\ref{fig:colors} and \ref{fig:colors2}).  For a limiting magnitude M$_{\rm V}$ = 25(24)[21], this corresponds to a maximum detection (luminosity) distance of $D_{\rm L} =$ 1070(680)[170]$M_{\rm ej,-2}^{1/4}$ Mpc and a co-moving volume $V = 2.8(0.9)[2\times 10^{-2}]$$M_{\rm ej,-2}^{3/4}$ Gpc$^{3}$.  The Palomar Transient Factory (PTF) 5-day cadence survey \citep{Law+:2009}, which surveys an active area $\sim 2700$ deg$^{2}$ to a limiting AB magnitude of 21, should therefore detect $\sim 1.4R_{-4}M_{\rm ej,-2}^{3/4}$ yr$^{-1}$.  Thus, if NS-NS mergers occur at the upper end of present rate estimates ($\sim 10^{-4}$ yr$^{-1}$) and $M_{\rm ej} \approx 10^{-2}M_{\sun}$ is indeed representative, current surveys such as PTF should ``blindly'' detect $\sim 1$ merger per year.  We emphasize, however, that the total predicted rate of events and their luminosity function is sensitive to the distribution of ejecta masses $M_{\rm ej}$, which in principle could range from $\sim 0 - 0.1M_{\sun}$ given present uncertainties (\S\ref{sec:nrichejecta}).

Prospects for detection are much better with the Large Synoptic Survey Telescope (LSST), which will image the entire sky down to a limiting magnitude $\sim 24.5$ every 3-4 nights and should detect NS-NS merger events at a rate $\sim 2 \times 10^{3}R_{-4}M_{\rm ej,-2}^{3/4}$ yr$^{-1}$.  Note that LSST is expected to come on-line in 2015, roughly coincident with Advanced LIGO/Virgo, and together they have the potential to completely revolutionize our understanding of compact object mergers.
   
Other thermal transients are predicted to occur in Nature on $\sim$ day timescales, which could be confused with NS-NS/NS-BH mergers.  Examples include ``.Ia'' SNe due to unstable thermonuclear He flashes from white dwarf binaries \citep{Bildsten+:2007,Poznanski+:2010} and Nickel-rich outflows from the accretion-induced collapse of WDs (\citealt{Metzger+:2009b}; Darbha et al.~2010).  Such events may originate from a similar stellar population to NS-NS/NS-BH mergers.  However, one ``smoking gun'' feature of kilonovae from NS-NS/NS-BH mergers is the presence of optical absorption lines due to heavy neutron-rich elements (as in $r$-process enriched halo stars; e.g. \citealt{Sneden+:2003}), which may not be present in white dwarf systems (although some $r$-process nuclei may be produced in neutrino-heated winds in the case of AIC; e.g. \citealt{Dessart+:2006}).  Thus, NS-NS/NS-BH mergers may be distinguishable from other transients with rapid, deep spectroscopic observations.  We plan to explore more detailed calculations of these $r$-process spectral features in future work.

In reality, only limited information will initially be available to transient searches (e.g. photometric colors, at best).  Thus, the reddening of kilonovae at times $\gtrsim t_{\rm peak}$ (Fig.~\ref{fig:colors}), if indeed robust, may be crucial for identifying these events.  In fact, of the variable sources from the Sloan Digital Sky Survey characterized by \citet{Sesar+:2007}, only a small fraction are as red as we predict kilonovae from NS-NS/NS-BH mergers to be at peak light (U$-$V $\gtrsim$ 2 following peak brightness; compare our results in Fig.~$\ref{fig:colors}$ with Fig.~4 of \citealt{Sesar+:2007}).  This suggests that a promising search strategy for detecting kilonovae is to trigger on anomalously red events of duration $\sim$ hour$-$day for more detailed follow-up observations.  

\subsubsection{Implications for the Origin of $r$-Process Elements}
\label{sec:rprocess}

%%%%%%%%%%%%%%%%%%%%%%%%%%%%%%%%%%%%%% FIG 3 %%%%%%%%%%%%%%%%%%%%%%%%%%%%%%
\begin{figure}
\resizebox{\hsize}{!}{\includegraphics[]{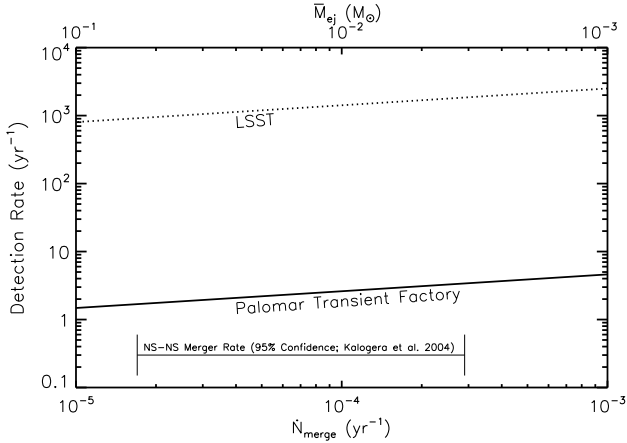}}
\caption{Rates of detected kilonovae from NS-NS/NS-BH mergers with present and upcoming surveys as a function of the merger rate $\dot{N}_{\rm merge}$ ({\it bottom axis}) or the average ejecta mass $\bar{M}_{\rm ej}$ ({\it top axis}), calculated under the assumption that NS-NS/NS-BH mergers are the primary source of $r$-process elements in the Galaxy.  Also shown are the NS-NS merger rate estimates (95$\%$ confidence interval) from Kalogera et al.~(2004).  }
\label{fig:detectionrates}
\end{figure}
%%%%%%%%%%%%%%%%%%%%%%%%%%%%%%%%%%%%%%%%%%%%%%%%%%%%%%%%%%%%%%%%%%%%%%%%%%%%
%%%%%%%%%%%%%%%%%%%%%%%%%%%%%%%%%%%%%%%%%%%%%%%%%%%%%%%%%%%%%%%%%%%%%%%%%%%%

The astrophysical origin of the $r$-process elements remains one of the great mysteries in nuclear astrophysics (see \citealt{Qian.Wasserberg:2007} for a recent review), with the two chief candidates being core-collapse supernovae (e.g. \citealt{Meyer+:1992}) and NS-NS/NS-BH mergers (\citealt{Freiburghaus+:1999}; see Fig.~\ref{fig:abundances}).  Since the luminosity of merger transients is directly related to their nucleosynthetic yield (eq.~[\ref{eq:lpeak}]), this implies that the detection of, or constraints on the rate of, kilonovae from NS mergers directly probes the origin of $r$-process elements.

As a concrete example, if one assumes that NS mergers are the dominant source of $r$-process elements in our Galaxy, then the mean mass ejected per event $\bar{M}_{\rm ej}$ and merger rate $\dot{N}_{\rm merge}$ are related by $\dot{N}_{\rm merge} = 10^{-4}{\,\rm yr^{-1}}(\bar{M}_{\rm ej}/10^{-2}M_{\sun})^{-1}$, where we have assumed a Galactic $r$-process production rate of $10^{-6}M_{\sun}$ yr$^{-1}$ (e.g. Qian 2000).  Since the kilonova luminosity $L_{\rm peak} \propto M_{\rm ej}^{1/2}$, the rate of detected transients $\propto L^{3/2}\dot{N}_{\rm merge} \propto \bar{M}_{\rm ej}^{-1/4} \propto \dot{N}_{\rm merge}^{1/4}$.  Figure \ref{fig:detectionrates} shows the expected detection rates versus $\dot{N}_{\rm merge}$ (or, equivalently, $\bar{M}_{\rm ej}$) with present and upcoming transient surveys if one assumes that NS-NS/NS-BH mergers are the dominant $r$-process source.  Note that within the current uncertainties in $\dot{N}_{\rm merge}$, {\it current transient surveys should detect a few events per year if mergers are indeed the dominant source of the $r$-process}, independent of $\bar{M}_{\rm ej}$.  In reality, the detection efficiency for low $\bar{M}_{\rm ej}$ may be somewhat lower than this simple estimate due to the shorter transient duration $t_{\rm peak} \propto M_{\rm ej}^{1/2}$ (eq.~[\ref{eq:tpeak}]).  

\section{Conclusions}
\label{sec:conclusions}

In their seminal paper on nucleosynthesis, \citet[B$^{2}$FH]{Burbidge+:1957} proposed that SNe are powered by the radioactive decay of $^{254}$Cf (cf.~\citealt{Burbidge+:1956}).  Although we now appreciate that most supernovae are powered by $^{56}$Ni and $^{56}$Co, the B$^{2}$FH picture of ``$r$-process-powered'' SNe still holds relevance for the neutron-rich ejecta from NS-NS/NS-BH mergers.  

In this paper we have presented the first calculations of the radioactively-powered transients from NS-NS/NS-BH mergers that self-consistently determine the radioactive heating using a nuclear reaction network and which accurately model the light curve and color evolution with a radiative transfer calculation.  Our main conclusions are summarized as follows:
\begin{itemize}
\item{The radioactive heating $\dot{E}$ on timescales $t_{\rm peak} \sim$ hours$-$days results [in approximately equal parts] from the fission and $\beta-$decays of heavy nuclei, which are produced by the $r$-process at much earlier times ($t \lesssim 1$ second; see Fig.~\ref{fig:edot}).  Our results for $\dot{E}$ at $t \sim t_{\rm peak}$ are relatively insensitive to the precise electron fraction and early-time expansion of the ejecta (e.g. whether it is dynamically-ejected or wind-driven), and to details of uncertain nuclear physics such as the theoretical nuclear mass model.}
\item{The net heating rate decreases approximately as a power law $\dot{Q} \propto t^{-\alpha}$ with $\alpha \sim 1.1-1.4$ for $t \sim $ hours$-$days, similar to the assumption $\dot{Q} \propto t^{-1}$ in the LP98 model.  The total heating rate is $\sim 3\times 10^{10}$ ergs s$^{-1}$ g$^{-1}$ at $t \approx 1$ day.  By calibrating the LP98 model using our results, we find an effective ``f'' parameter $\sim 3\times 10^{-6}$ which is generally much lower than previously assumed.}
\item{$\beta-$decay electrons ($\S\ref{sec:betadecays}$) and fission daughter nuclei ($\S\ref{sec:fission}$) both thermalize with the plasma on timescales $\sim t_{\rm peak}$, while only a portion of the $\gamma-$rays likely thermalize.  We estimate that the {\it net} thermalization efficiency is $\epsilon_{\rm therm} \sim 0.5-1$ ($\S\ref{sec:therm}$).}
\item{For an ejecta mass $M_{\rm ej} \sim 10^{-2}M_{\sun}$ to $10^{-3}M_{\sun}$ we predict a transient that peaks on a timescale $\sim 1$ day at a bolometric and V-band luminosity $\sim 10^{42}$ ergs s$^{-1}$ ($M_{\rm B}$ = $-$16) and $\sim 3\times 10^{41}$ ergs s$^{-1}$ ($M_{\rm V}$ = $-$15), respectively (Figs.~\ref{fig:lc} and \ref{fig:colors}).}
\item{We argue that the transition metal $r$-process elements are likely to have UV absorption due to line blanketing (like Fe peak nuclei).  As a result, we predict that NS merger transients will be relatively red (and redden in time; see Figs.~\ref{fig:colors} and \ref{fig:colors2}), a prediction not captured by assuming single-temperature blackbody emission.  More detailed models of the color evolution of NS merger transients will require a better understanding of the UV and IR spectral lines of second and third $r$-process peak elements ($\S\ref{sec:sedona}$).  The presence of absorption lines due to heavy $r$-process elements is one ``smoking gun'' prediction of NS merger transients.}
\item{Because NS merger transients are isotropic, they can in principle be detected in three independent ways: in coincidence with a detected GW source; following a short-duration GRB; and with blind optical/NIR transient surveys.}
\item{Given the low luminosities and rapid evolution of kilonovae from NS mergers, their detection will require close collaboration between the GW and astronomical communities.  Given the unique observational signature of kilonovae, the real-time follow-up of GW detections with sensitive, wide-field telescopes could improve the effective sensitivity of LIGO/Virgo.}
\item{For an average ejecta mass $\bar{M}_{\rm ej} \approx 10^{-2}M_{\sun}$, current surveys such as PTF should ``blindly'' detect $\sim 1$ NS merger transient per year if the merger rates lies at the high end of present estimates; LSST should detect $\sim$ one thousand per year under the same assumptions.}
\item{Since the luminosity and detection rate of NS merger transients is closely related to the yield of heavy neutron-rich elements, {\it current} transient surveys are directly probing the unknown astrophysical origin of the $r$-process (Fig.~\ref{fig:detectionrates}).  Holding the total $r$-process injection rate in the Milky Way fixed at $10^{-6}M_{\sun}$ yr$^{-1}$ implies a detection rate $\sim$ few yr$^{-1}$ and $\sim 10^{3}$ yr$^{-1}$ for PTF and LSST, respectively, independent of the average ejecta mass.}
\end{itemize}

\section*{Acknowledgments}

We thank G.~Wahlgren, P.~Shawhan, C.~Sneden, F.-K.~Thielemann, and C.~Blake for
helpful conversations and useful information.  We thank V.~Petrosian for suggesting the term ``kilonovae'' to describe NS merger transients.  Support for BDM was
provided by NASA through an Einstein Fellowship (SAO \#PF9-00065; NASA
prime award number \#NAS8-03060).  AA and GMP are partly supported by the
  Deutsche Forschungsgemeinschaft through contract SFB 634 and the
  Helmholtz Alliance \emph{Cosmic Matter in the Laboratory}.  EQ was supported in part by the
Miller Institute for Basic Research in Science, University of
California Berkeley, and by the David and Lucile Packard Foundation.
Support for DK was provided by NASA through Hubble fellowship grant
\#HST-HF-01208.01-A awarded by the Space Telescope Science Institute,
which is operated by the Association of Universities for Research in
Astronomy, Inc., for NASA, under contract NAS 5-26555.  This research
has been supported in part by the DOE SciDAC Program
(DE-FC02-06ER41438).  Support for RT and PN was provided by the
Director, Office of Science, Office of High Energy Physics, of the
U.S. Department of Energy under Contract No. DE-AC02-05CH11231.  IP was supported in part by SCOPES project  No.~IZ73Z0-128180/1 awarded by the Swiss National Science Foundation, and by  Russia Ministry of education and science, contract number 02.740.11.0250.

%%%%%%%%%%%%%%%%%%%%%%%%%%%%%%%%%%%%%%%%%%%%%%%%%%%%%%%%%%%%%%%%%%
%\bibliographystyle{mn2e}
%\bibliography{../biblio/bibliography}
%\bibliography{ms}

\label{lastpage}

\end{document}